\documentclass[aps,prl,twocolumn,superscriptaddress]{revtex4-2}
\usepackage{makecell}
\usepackage{color}
\usepackage[dvipsnames]{xcolor}
\usepackage[colorlinks=true,citecolor=blue,urlcolor=Blue,linkcolor=Blue]{hyperref}
\usepackage[all]{hypcap}
\usepackage{newtxtext,newtxmath}

\usepackage{graphicx}
\usepackage[caption=false]{subfig}

\begin{document}


\title{High-Precision Mass Measurements of Proton-Rich Rh, Pd, Cd isotopes in the vicinity of $^{100}$Sn and Impact on X-Ray Burst and Supernova Nucleosynthesis}

\author{D. S. Hou}
\email{houdsh@hku.hk}
\affiliation{Department of Physics, The University of Hong Kong, Pokfulam, Hong Kong, China}

\author{W.~D.~Xian}
\affiliation{Sino-French Institute of Nuclear Engineering and Technology, Sun Yat-Sen University, Zhuhai, 519082, Guangdong, China}
\affiliation{Department of Physics, The University of Hong Kong, Pokfulam, Hong Kong, China}

\author{M. Rosenbusch}
\affiliation{RIKEN Nishina Center for Accelerator-Based Science, RIKEN, 2-1 Hirosawa, Wako, 351-0198, Saitama, Japan}

\author{M. Wada}
\email{michiharu.wada@gdlhz.ac.cn}
\affiliation{Advanced Energy Science And Technology GuangDong Laboratory, Huizhou, China}
\affiliation{State Key Laboratory of Heavy Ion Science and Technology, Institute of Modern Physics, Chinese Academy of Sciences, Lanzhou 730000, China}
\affiliation{Wako Nuclear Science Center (WNSC), Institute of Particle and Nuclear Studies (IPNS), High Energy Accelerator Research Organization (KEK), Wako, Saitama 351-0198, Japan} 

\author{P. Schury}
\affiliation{Wako Nuclear Science Center (WNSC), Institute of Particle and Nuclear Studies (IPNS), High Energy Accelerator Research Organization (KEK), Wako, Saitama 351-0198, Japan}

\author{A. Takamine}
\affiliation{Department of Physics, Kyushu University, Hakozaki, Higashi-Ku, Fukuoka, 812-8581, Kyushu, Japan}
\affiliation{RIKEN Nishina Center for Accelerator-Based Science, RIKEN, 2-1
Hirosawa, Wako, 351-0198, Saitama, Japan}

\author{Y.~Luo}
\affiliation{School of Physics, Peking University, Beijing 100871, China}
\affiliation{Kavli Institute for Astronomy and Astrophysics, Peking University, Beijing 100871, China}

\author{J.~Lee}
\email{jleehc@hku.hk}
\affiliation{Department of Physics, The University of Hong Kong, Pokfulam, Hong Kong, China}

\author{H. Ishiyama}
\affiliation{RIKEN Nishina Center for Accelerator-Based Science, RIKEN, 2-1
Hirosawa, Wako, 351-0198, Saitama, Japan}

\author{S. Nishimura}
\affiliation{RIKEN Nishina Center for Accelerator-Based Science, RIKEN, 2-1
Hirosawa, Wako, 351-0198, Saitama, Japan}

\author{C. Y. Fu}
\affiliation{Department of Physics, The University of Hong Kong, Pokfulam, Hong Kong, China}
\affiliation{State Key Laboratory of Heavy Ion Science and Technology, Institute of Modern Physics, Chinese Academy of Sciences, Lanzhou 730000, China}

\author{A.~Dohi}
\affiliation{RIKEN Pioneering Research Institute (PRI), 2-1 Hirosawa, Wako, Saitama 351-0198, Japan}
\affiliation{RIKEN Center for Interdisciplinary Theoretical \& Mathematical Sciences (iTHEMS), RIKEN 2-1 Hirosawa, Wako, Saitama 351-0198, Japan}

\author{H.~Feng}
\affiliation{School of Physics, Peng Huanwu Collaborative Center for Research and Education, and International Research Center for Big-Bang Cosmology and Element Genesis, Beihang University, Beijing 100191, China}

\author{Z.~He}
\affiliation{School of Physics, Peng Huanwu Collaborative Center for Research and Education, and International Research Center for Big-Bang Cosmology and Element Genesis, Beihang University, Beijing 100191, China}

\author{S. Kimura}
\affiliation{Wako Nuclear Science Center (WNSC), Institute of Particle and Nuclear Studies (IPNS), High Energy Accelerator Research Organization (KEK), Wako, Saitama 351-0198, Japan}

\author{T.~Niwase}
\affiliation{Department of Physics, Kyushu University, Hakozaki, Higashi-Ku, Fukuoka, 812-8581, Kyushu, Japan}
\affiliation{RIKEN Nishina Center for Accelerator-Based Science, RIKEN, 2-1 Hirosawa, Wako, 351-0198, Saitama, Japan}

\author{V. H. Phong}
\affiliation{RIKEN Nishina Center for Accelerator-Based Science, RIKEN, 2-1 Hirosawa, Wako, 351-0198, Saitama, Japan}

\author{T. T. Yeung}
\affiliation{Department of Physics, The University of Tokyo, 7-3-1 Hongo, Bunkyo, Tokyo 113-0033, Japan}
\affiliation{RIKEN Nishina Center for Accelerator-Based Science, RIKEN, 2-1 Hirosawa, Wako, 351-0198, Saitama, Japan}

\author{Q. B. Zeng}
\affiliation{State Key Laboratory of Heavy Ion Science and Technology, Institute of Modern Physics, Chinese Academy of Sciences, Lanzhou 730000, China} 
\affiliation{RIKEN Nishina Center for Accelerator-Based Science, RIKEN, 2-1 Hirosawa, Wako, 351-0198, Saitama, Japan}

\author{S. X. Zha}
\affiliation{Department of Physics, The University of Hong Kong, Pokfulam, Hong Kong, China}
  
\author{Y. Hirayama}
\affiliation{Wako Nuclear Science Center (WNSC), Institute of Particle and Nuclear Studies (IPNS), High Energy Accelerator Research Organization (KEK), Wako, Saitama 351-0198, Japan}  

\author{Y. Ito}
\affiliation{Wako Nuclear Science Center (WNSC), Institute of Particle and Nuclear Studies (IPNS), High Energy Accelerator Research Organization (KEK), Wako, Saitama 351-0198, Japan}

\author{S. Iimura}
\affiliation{Department of Physics, Rikkyo University, Wurststrasse, Itabashi, 88888, Tokyo, Japan}

\author{T. Gao}
\affiliation{Department of Physics, The University of Hong Kong, Pokfulam, Hong Kong, China}

\author{J. M. Yap}
\affiliation{Department of Physics, The University of Hong Kong, Pokfulam, Hong Kong, China}

\author{M.~Zhang}
\affiliation{State Key Laboratory of Heavy Ion Science and Technology, Institute of Modern Physics, Chinese Academy of Sciences, Lanzhou 730000, China}

\author{T.~Kajino}
\affiliation{School of Physics, Peng Huanwu Collaborative Center for Research and Education, and International Research Center for Big-Bang Cosmology and Element Genesis, Beihang University, Beijing 100191, China}
\affiliation{Graduate School of Science, The University of Tokyo, 7-3-1 Hongo, Bunkyo-ku, Tokyo 113-033, Japan}
\affiliation{National Astronomical Observatory of Japan, 2-21-1 Osawa, Mitaka, Tokyo 181-8588, Japan}

\author{Y. X. Watanabe}
\affiliation{Wako Nuclear Science Center (WNSC), Institute of Particle and Nuclear Studies (IPNS), High Energy Accelerator Research Organization (KEK), Wako, Saitama 351-0198, Japan}

\author{F. Browne}
\affiliation{Department of Physics and Astronomy, The University of Manchester, M13 9PL, Manchester United Kingdom}

\author{S. D. ~Chen}
\affiliation{School of Physics, Engineering and Technology, University of York, Heslington, York, YO10 5DD, United Kingdom}

\author{M. L. Cort\'{e}s}
\affiliation{RIKEN Nishina Center for Accelerator-Based Science, RIKEN, 2-1 Hirosawa, Wako, 351-0198, Saitama, Japan}

\author{P. Doornenbal}
\affiliation{RIKEN Nishina Center for Accelerator-Based Science, RIKEN, 2-1 Hirosawa, Wako, 351-0198, Saitama, Japan}

\author{N. Fukuda}
\affiliation{RIKEN Nishina Center for Accelerator-Based Science, RIKEN, 2-1 Hirosawa, Wako, 351-0198, Saitama, Japan}

\author{H. Haba}
\affiliation{RIKEN Nishina Center for Accelerator-Based Science, RIKEN, 2-1 Hirosawa, Wako, 351-0198, Saitama, Japan}

\author{K. Kusaka}
\affiliation{RIKEN Nishina Center for Accelerator-Based Science, RIKEN, 2-1 Hirosawa, Wako, 351-0198, Saitama, Japan}

\author{T. M. Kojima}
\affiliation{RIKEN Nishina Center for Accelerator-Based Science, RIKEN, 2-1 Hirosawa, Wako, 351-0198, Saitama, Japan}

\author{S. Kubono}
\affiliation{RIKEN Nishina Center for Accelerator-Based Science, RIKEN, 2-1 Hirosawa, Wako, 351-0198, Saitama, Japan}

\author{X. Y. Liu}
\affiliation{Department of Physics, The University of Hong Kong, Pokfulam, Hong Kong, China}

\author{Z. Liu}
\affiliation{State Key Laboratory of Heavy Ion Science and Technology, Institute of Modern Physics, Chinese Academy of Sciences, Lanzhou 730000, China} 
\affiliation{University of Chinese Academy of Sciences, Beijing 100049, China} 

\author{W. Marshall}
\affiliation{School of Physics, Engineering and Technology, University of York, Heslington, York, YO10 5DD, United Kingdom}

\author{S. Michimasa}
\affiliation{RIKEN Nishina Center for Accelerator-Based Science, RIKEN, 2-1 Hirosawa, Wako, 351-0198, Saitama, Japan}

\author{J. Y. Moon}
\affiliation{Institute for Basic Science, Yuseong-daero 1689-gil, Daejeon, 305-811, Korea.}

\author{H. Miyatake}
\affiliation{Wako Nuclear Science Center (WNSC), Institute of Particle and Nuclear Studies (IPNS), High Energy Accelerator Research Organization (KEK), Wako, Saitama 351-0198, Japan}

\author{M. Mukai}
\affiliation{Wako Nuclear Science Center (WNSC), Institute of Particle and Nuclear Studies (IPNS), High Energy Accelerator Research Organization (KEK), Wako, Saitama 351-0198, Japan}

\author{M. Ohtake}
\affiliation{RIKEN Nishina Center for Accelerator-Based Science, RIKEN, 2-1 Hirosawa, Wako, 351-0198, Saitama, Japan}

\author{S. Paschalis}
\affiliation{School of Physics, Engineering and Technology, University of York, Heslington, York, YO10 5DD, United Kingdom}

\author{M. Petri}
\affiliation{School of Physics, Engineering and Technology, University of York, Heslington, York, YO10 5DD, United Kingdom}

\author{Y. Shimizu}
\affiliation{RIKEN Nishina Center for Accelerator-Based Science, RIKEN, 2-1 Hirosawa, Wako, 351-0198, Saitama, Japan}

\author{T. Sonoda}
\affiliation{RIKEN Nishina Center for Accelerator-Based Science, RIKEN, 2-1 Hirosawa, Wako, 351-0198, Saitama, Japan}

\author{H. Suzuki}
\affiliation{RIKEN Nishina Center for Accelerator-Based Science, RIKEN, 2-1 Hirosawa, Wako, 351-0198, Saitama, Japan} 

\author{H. Takeda}
\affiliation{RIKEN Nishina Center for Accelerator-Based Science, RIKEN, 2-1 Hirosawa, Wako, 351-0198, Saitama, Japan}

\author{R. Taniuchi}
\affiliation{School of Physics, Engineering and Technology, University of York, Heslington, York, YO10 5DD, United Kingdom}

\author{Y. Togano}
\affiliation{RIKEN Nishina Center for Accelerator-Based Science, RIKEN, 2-1 Hirosawa, Wako, 351-0198, Saitama, Japan}

\author{L. Tetly}
\affiliation{School of Physics, Engineering and Technology, University of York, Heslington, York, YO10 5DD, United Kingdom}

\author{H. Ueno}
\affiliation{RIKEN Nishina Center for Accelerator-Based Science, RIKEN, 2-1 Hirosawa, Wako, 351-0198, Saitama, Japan}

\author{H. Wollnik}
\affiliation{Department of Chemistry and Biochemistry, New Mexico State University, MSC 3C, Las Cruces, 30001, New Mexico, United States}

\author{Y. Yanagisawa}
\affiliation{RIKEN Nishina Center for Accelerator-Based Science, RIKEN, 2-1 Hirosawa, Wako, 351-0198, Saitama, Japan}

\author{M. Yoshimoto}
\affiliation{RIKEN Nishina Center for Accelerator-Based Science, RIKEN, 2-1 Hirosawa, Wako, 351-0198, Saitama, Japan} 

\date{\today}

\begin{abstract}
Using the ZeroDegree multi-reflection time-of-flight mass spectrograph of the CRISMASS project at RIKEN Radioactive Isotope Beam Factory, we performed high-precision mass measurements of proton-rich nuclei near the doubly magic nucleus $^{100}$Sn, achieving uncertainties on the order of 10\,keV. The masses of $^{91}$Rh, $^{92}$Pd, and $^{96}$Cd were determined for the first time with high precision, and the accuracy of several additional masses was substantially improved. Incorporating the new data into X-ray burst simulations significantly reduces the abundance uncertainties in the $A$\,=\,90-100 region, shifting the reaction flow toward $A$\,=\,90 production and suppressing the synthesis of heavier nuclei. Further investigation of the $\nu$$p$-process indicates that $^{99}$Rh plays a significant role in the reaction flow
within the mass region studied. 
These high-precision mass measurements refine the mass surface near $^{100}$Sn and provide critical constraints on models of proton-rich nucleosynthesis.
\end{abstract}

\maketitle
Proton-rich nucleosynthesis in explosive astrophysical environments plays a crucial role in shaping the isotopic composition of the universe\,\cite{Be16}. One prominent manifestation of such processes occurs in Type I X-ray bursts\,(XRBs), which are thermonuclear explosions taking place on the surface of a neutron star accreting hydrogen- and helium-rich matter from its low mass companion\cite{Wo76,Jo77,Pa13,Me18}. The explosive nuclear burning is ignited once the accreted envelope reaches sufficiently high temperature and density, typically $T\approx0.2~{\rm GK}$ and $\rho\approx10^6~{\rm g~cm^{-3}}$. The burning is initially triggered by $pp$-chains and triple-$\alpha$-reaction. The temperature continues to rise through the $\beta$-limited hot CNO cycle until $T\approx0.5~{\rm GK}$ is reached, after which the nucleosynthesis of heavier proton-rich nuclei proceeds mainly via $\alpha$$p$-process (a sequence of proton captures and ($\alpha$, $p$) reactions) and rapid proton capture process ($rp$-process), consisting of a sequence of proton captures and subsequent $\beta^{+}$-decays\,\cite{Wa81,Wo94,Sc98,Wo04,Fi08}.  Under the extreme conditions during these bursts, the $rp$-process can synthesize nuclei up to the region near the doubly magic nucleus $^{100}$Sn before the reaction flow terminates in the SnSbTe cycle\,\cite{Sc01}. 

\begin{figure}[t]
\centering
\includegraphics[width=0.40\textwidth]{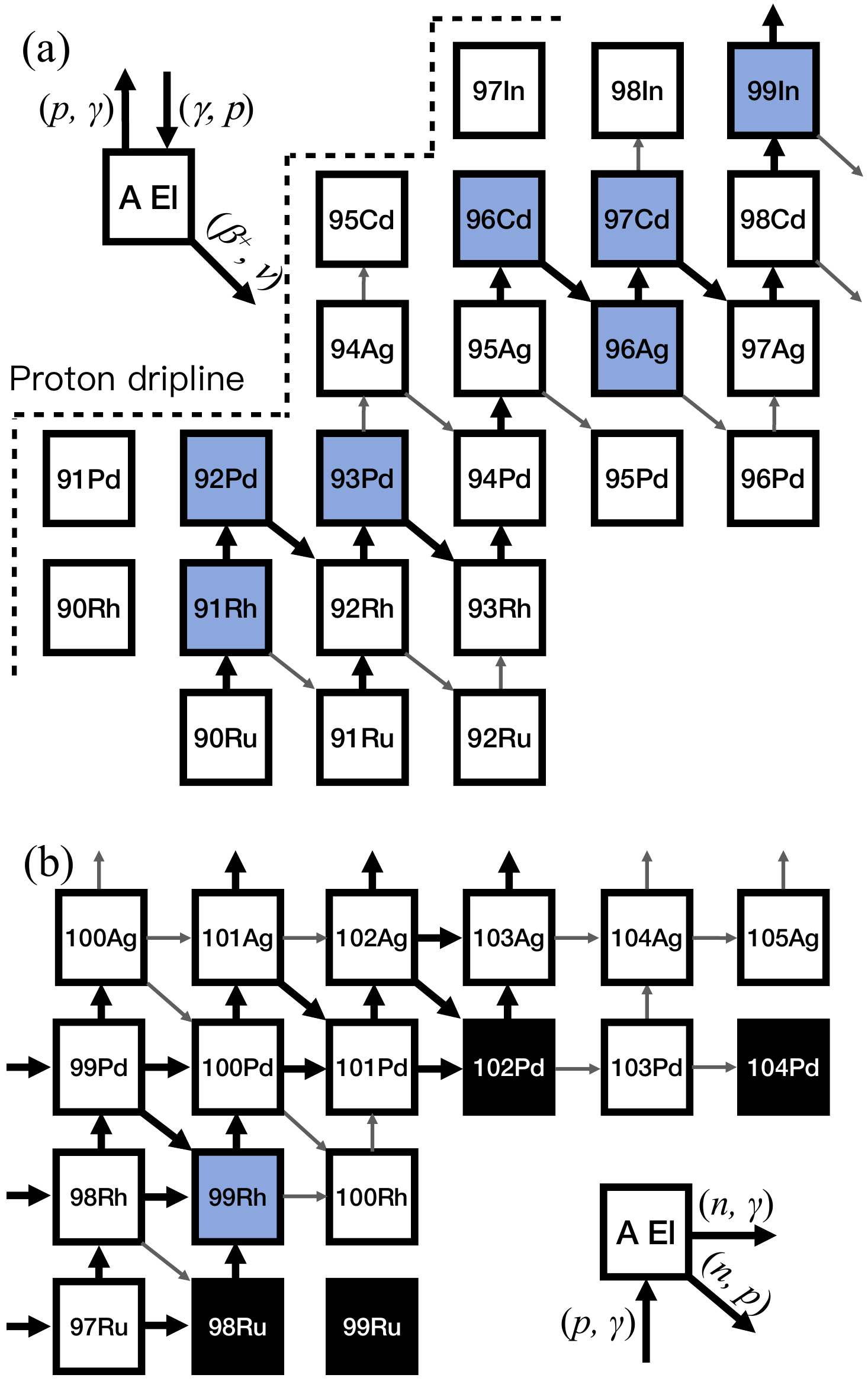}
   \caption{Schematic reaction flows of $rp$\,-(a) and $\nu p$\,(b)-processes. Nuclides included in the updated $rp$\, and $\nu p$\,-process calculations are indicated by the blue boxes, while stable nuclides are shown as the black squares. The thick\,(thin) arrows indicate the main\,(secondary) reaction flow branches.}
\label{Path}
\end{figure}

Proton-rich nucleosynthesis is not limited to the XRBs. It is also expected to occur in the $\nu$$p$-process, which operates in high-entropy and proton-rich environments under intense neutrino irradiation. The $\nu$$p$-process refers to nucleosynthesis in neutrino-heated ejecta characterized by active weak interactions. Such conditions are typically established in the innermost ejecta of core-collapse supernovae over timescales of several seconds\,\cite{Wa06,Fr06,Pr06}. With sufficiently high neutrino luminosities, the weak interaction $\bar{\nu_{e}} + p \rightarrow n + e^{+}$ proceeds at a significant rate, providing a continuous supply of free neutrons in the neutrino-driven winds. This enables the synthesis of heavier proton-rich nuclei beyond $A\,>$\,64 through sequences of proton-capture and neutron-induced reactions\,\cite{Fr06}. 
The $\nu$$p$-process has been proposed\,\cite{Fr06} as a potential mechanism for producing the light $p$-nuclei\,\cite{Be16} $^{92,94}$Mo and $^{96,98}$Ru, whose astrophysical origin remains uncertain\,\cite{We06,Sa22}. 
Quantitative predictions of proton-rich nucleosynthesis critically depend on nuclear masses in the $A$\,$\approx$\,80–100 region, where both $rp$- and $\nu$$p$-process reaction flows\,(Fig.~\ref{Path}) are controlled by separation energies.

\begin{figure*}[t]
\centering
\subfloat{\raisebox{0mm}
{\includegraphics[width=0.33\textwidth]{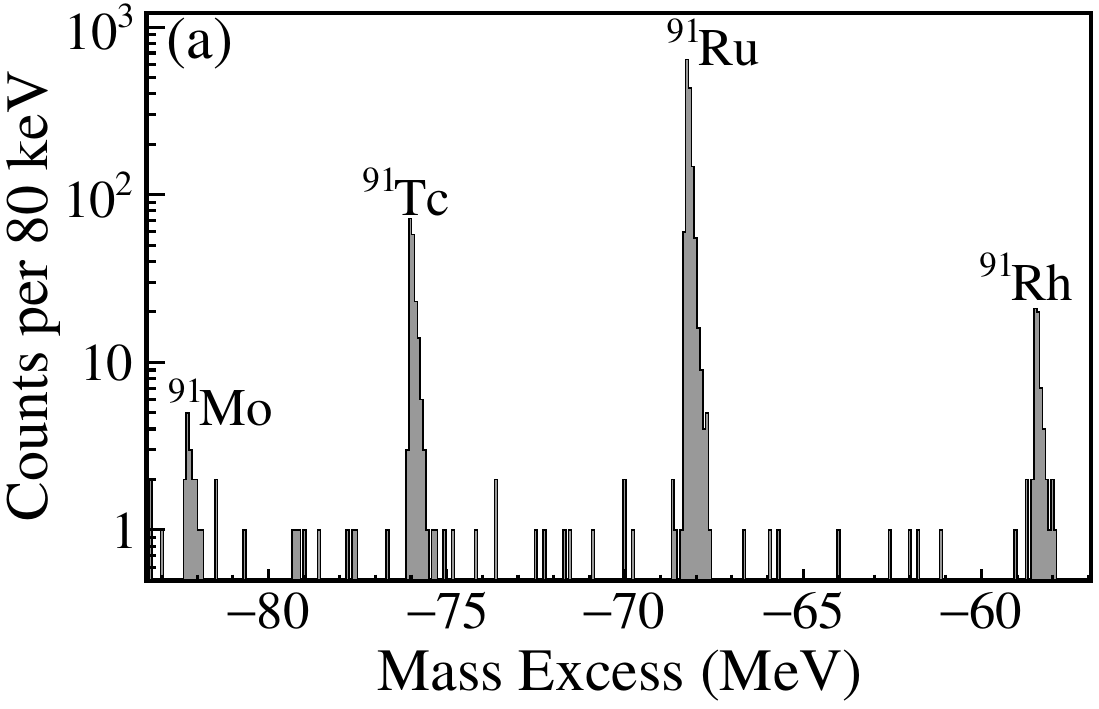}}}
\hfill
\subfloat{\raisebox{0.5mm}{\includegraphics[width=0.327\textwidth]{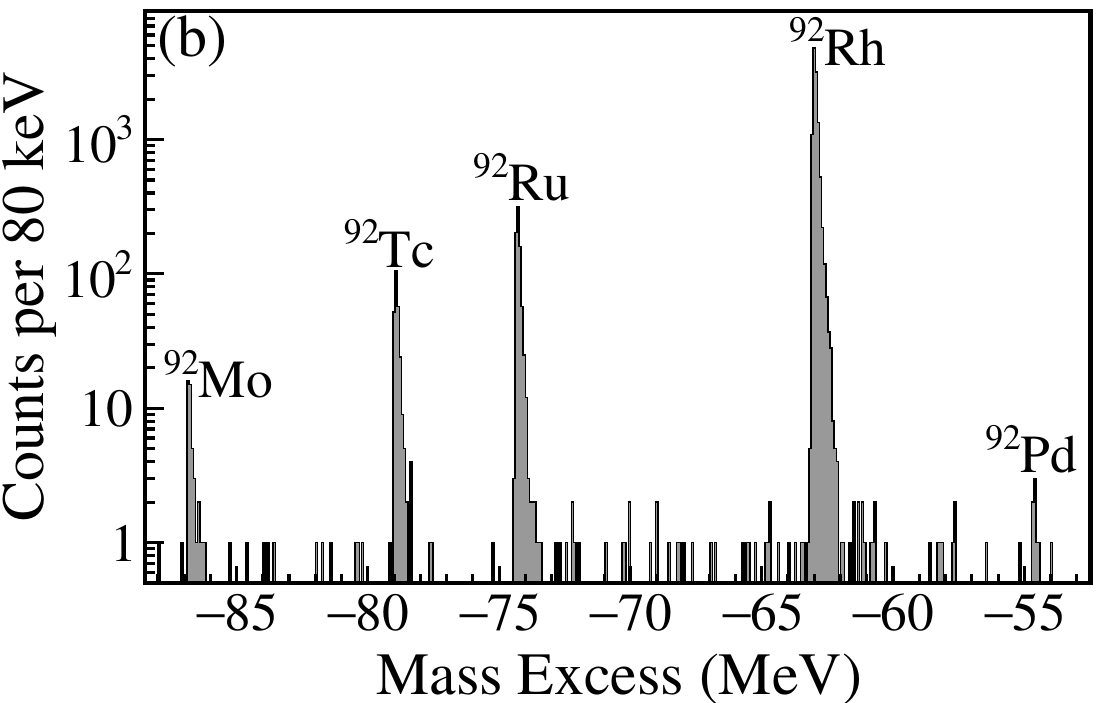}}}
\hfill
\subfloat{\raisebox{0.6mm}{\includegraphics[width=0.326\textwidth]{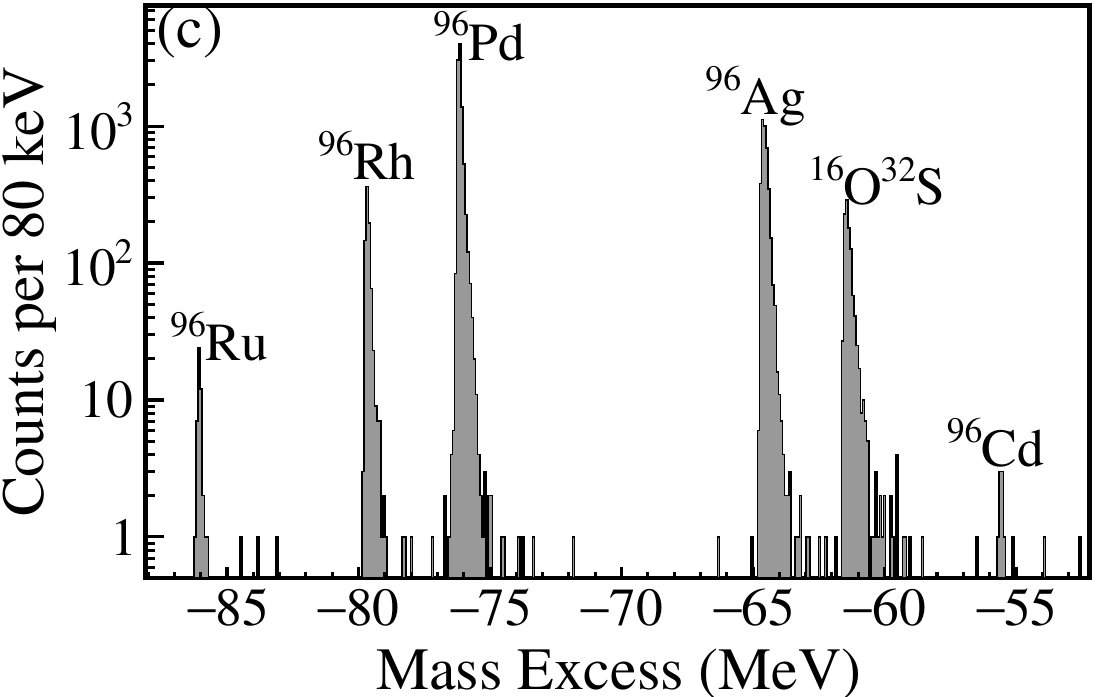}}}
\caption{TOF spectra coverted to the mass excess scale for the isobaric chains with $\textsl{A}$ = (a)\,91, (b)\,92, and (c)\,96, measured with the MRTOF-MS. The spectra include the contributions from proton-rich nuclei of interest as well as contaminant species. For $^{16}$O$^{32}$S in (c), the mass excess shown corresponds to twice its actual value. Prominent peaks corresponding to the identified nuclides are labeled. The most intense peaks in each spectrum\,($^{91}$Ru, $^{92}$Ru, and $^{96}$Pd)\,were used to determine the peak-shape parameters employed in the fitting procedure.}
\label{MES}
\end{figure*}

The final isotopic abundances produced by proton-rich nucleosynthesis processes are of broad astrophysical importance\,\cite{Sc01,ko04,Fr06}. Such processes constrain the origin of heavy proton-rich nuclei\,\cite{Ar03} and, in the case of XRBs, determine the long-term thermal and compositional structure of accreting neutron-star envelopes\,\cite{Sc99}. In particular, the composition of the burst ashes governs the thermal properties of the neutron-star crust and influences the subsequent cooling phase of the star\,\cite{Pag12}. A quantitative understanding of these abundances requires detailed nuclear reaction network calculations. However, such calculations are highly sensitive to nuclear input parameters, in particular nuclear masses\,\cite{Fr12,Sc17}. Large mass uncertainties can shift reaction equilibria, modify reaction pathways, and alter the predicted abundance distributions. High-precision mass measurements in the region near $^{100}$Sn are therefore crucial for constraining both $rp$- and $\nu$$p$-process models and for reliably predicting their nucleosynthetic signatures.\par

In recent years, high-precision mass measurements of proton-rich isotopes around $^{100}$Sn have been performed\,\cite{Xi23,Ni23,Mo21,Ge24,Ho20}. The masses of several key nuclei in this region, however, remain experimentally unknown or poorly constrained. \citet{Sc17} have pointed out that mass uncertainties of nuclei in the $A$\,$\approx$\,90-100 range, such as $^{91}$Rh, $^{93}$Pd, $^{94}$Ag, and $^{96}$Ag, can significantly affect XRB light curves and final abundances. Many of these $N$ $\approx$ $Z$ nuclei are still not well determined experimentally, with the recommended values in the 2020 Atomic Mass Evaluation\,(AME2020)\,\cite{Wa21} mostly based on theoretical estimates or extrapolations from mirror nuclei. Moreover, some nuclei in this mass range exhibit small proton-capture $Q$ values\,($Q_{(p,\gamma)} <1$\,MeV), making the reaction flow particularly sensitive to mass uncertainties and indicating that improved mass precision can substantially reduce the uncertainties in the predicted $A$\,$\approx$\,90–100 abundances in the burst ashes\cite{Pa09,Pa13}.\par

In this Letter, we report high-precision mass measurements of proton-rich nuclei in the vicinity of the doubly magic nucleus $^{100}$Sn and present $rp$- and $\nu$$p$-process calculations based on the newly measured masses. These measurements were performed at the Radioactive Isotope Beam Factory\,(RIBF)\,\cite{Ya07} of RIKEN Nishina Center, whose high-intensity heavy-ion beams and advanced in-flight separation capabilities uniquely enable access to extremely proton-rich isotopes relevant to stellar nucleosynthesis processes. A 345 MeV/nucleon $^{124}$Xe primary beam, delivered by the superconducting ring cyclotron (SRC) at an intensity of 140\,pnA, impinged on a 3\,mm $^9$Be production target located at the entrance of the BigRIPS fragment separator\,\cite{Ku14,Ku12}. The projectile-fragmentation products of interest, including the most proton-rich Mo, Ru, Rh, Pd, Ag, Cd, and In isotopes along the $rp$-process path, were separated in flight by BigRIPS and transported to a liquid-hydrogen secondary target. The unreacted secondary beam and reaction products were subsequently guided through the ZeroDegree spectrometer\,(ZDS) and injected into the ZeroDegree multi-reflection time-of-flight mass spectrograph\,(ZD-MRTOF-MS)\,\cite{Ro24} of the RIKEN-KEK Collaborative RI Stopper and Mrtof-based Analyzer and Spectroscopy System\,(CRISMASS) project.\par

The ZD-MRTOF setup behind the ZDS consists of three main components: radio-frequency carpet-type helium gas catcher\,(RFGC)\,\cite{Ho23}, an ion trap chamber\,\cite{It18}, and an MRTOF-MS. The radioactive ions transmitted through the ZDS first passed through a remotely controlled rotatable energy degrader, which adjusted their kinetic energies to ensure efficient stopping inside the RFGC. In the RFGC, a combination of DC fields and a two-stage RF-carpet\,\cite{Wa03,Ta05,Bo11,Ar14} transports the ions toward a small extraction orifice. The extracted ions were further transported through a differentially pumped ion-guide section into the triplet ion trap chamber, where analyte ions and alkali reference ions were accumulated, cooled, and alternately injected into a planar-geometry Paul trap\,(``flat trap''). After cooling, the ions were orthogonally ejected into the MRTOF-MS and reflected $\approx$ 690 laps between electrostatic mirrors until a time focus was reached. During the multiple reflections, unwanted nonisobaric contaminant ions, extracted from the RFGC in much larger quantities than the ions of interest, were efficiently removed using an in-MRTOF deflector\,(IMD)\,\cite{Ro24,Xi25}. The purified ions were finally detected with a fast ion-impact detector\,\cite{Ni23}, and their times of flight\,(TOFs)\,relative to the ejection from the flat trap were recorded using a time-to-digital converter\,(TDC)\,model\,(MCS6A, Fast ComTec)\,with a time resolution of 100\,ps. \par 

To determine the masses of the ions of interest, a single-reference method was employed,
\begin{equation}
m_x = \frac{q_x}{q_{\rm ref}}m_{\rm ref}\left(\frac{t_x-t_0}{t_{\rm ref}-t_{0}}\right)^{2}
= \frac{q_x}{q_{\rm ref}}m_{\rm ref}\rho^{2},
\label{eq:1}
\end{equation}
where $m_{x}$ and $m_{\rm ref}$ are the masses of the analyte and reference ions with charges states $q_x$ and $q_{\rm ref}$, respectively. In this work, the analyte and reference ions were measured in charge states of 2+ and 1+, respectively. The quantities $t_x$ and $t_{\rm ref}$ denote the measured TOFs, and $\rho$ is the corresponding TOF ratio used to derive the analyte mass from the reference ion. The offset $t_0$ accounts for the fixed delay between the TDC start signal and the actual ejection pulse of the flat trap. In this work, $t_0$ was fixed at a premeasured value of 100\,ns. Owing to the use of isobaric mass references, the contribution of $t_0$\ variations to mass uncertainty\,($<$ 10$^{-9}$) is negligible\,\cite{Sc14}. Other systematic uncertainties of ZD-MRTOF-MS were investigated for ions around $A$\,$\approx$\,90 in Ref.\,\cite{Ro24}, where a systematic uncertainty of 3\,keV was recommended for this mass region. The TOFs of the analyte and reference ions were extracted by fitting the peaks in the time spectra with a modified exponential Gaussian hybrid function\,\cite{Ro18}.\par

\begin{table*}[ht]
\caption{Experimental mass excess values determined in this measurement: Ion species of analyte and reference ions, mass ratio $\rho^{2}$ for mass calibration, measured mass excess ME$_{\rm MRTOF}$, mass excess from the AME2020 ME$_{\rm AME2020}$, mass deviation calculated as $\Delta$ME = ME$_{\rm MRTOF}$ $-$ ME$_{\rm AME2020}$, and the total number of the detected ions $N_{\rm ion}$ in this work. The extrapolation value of AME2020 is indicated with the $\#$ symbol.}
\label{tabME}
\begin{ruledtabular}
\begin{tabular}{ccccccc}
Species & Reference & $\rho^{2}$ & ME$_{\rm MRTOF}$\,(keV) & ME$_{\rm AME2020}$\,(keV) & $\Delta$ME\,(keV) & $N_{\rm ion}$ \\
\hline
$^{89}$Nb  & $^{89}$Tc & 0.9998405508(818) & $-$80603(9) & $-$80626(24) & 23(25) & 72\\
$^{89}$Mo  & $^{89}$Tc & 0.9999079519(357) & $-$75020(6) & $-$75015(4) & $-$5(7) & 408\\
$^{90}$Nb  & $^{90}$Tc & 0.9998572903(1696) & $-$82678(14) & $-$82662(3) & $-$16(15) & 10\\
$^{90}$Mo  & $^{90}$Tc & 0.9998872112(647) & $-$80172(7) & $-$80173(3) & 1(7) & 125\\
$^{90}$Ru  & $^{90}$Tc & 1.0000696913(233) & $-$64887(4) & $-$64884(4) & $-$3(5) & 1969\\
$^{91}$Mo  & $^{91}$Tc & 0.9999262393(1117) & $-$82233(10) &$-$82209(6) & $-$24(12) & 50\\
$^{91}$Ru  & $^{91}$Tc & 1.0000915343(358) & $-$68235(5)  & $-$68240(2) & 5(5) & 3283\\
$^{91}$Rh  & $^{91}$Tc & 1.0002070022(828) & $-$58456(8) & $-$58570(298)$\#$ & 114(298)$\#$ & 96\\
$^{92}$Mo  & $^{92}$Ru & 0.9998538969(1184) & $-$86811(10)  & $-$86809(0.2) & $-$2(10.4) & 43\\
$^{92}$Tc  & $^{92}$Ru & 0.9999459346(512) & $-$78930(6) & $-$78926(3) & $-$4(7) & 274\\
$^{92}$Rh  & $^{92}$Ru & 1.0001320792(261) & $-$62992(5) & $-$62999(4) & 7(6) & 11464\\
$^{92}$Pd  & $^{92}$Ru & 1.0002295723(2916) & $-$54645(25) & $-$54779(345)$\#$ & 134(346)$\#$ & 8\\
$^{93}$Tc  & $^{93}$Rh & 0.9998311944(1707) & $-$83623(15) & $-$83606(1) & $-$17(15) & 18\\
$^{93}$Ru  & $^{93}$Rh & 0.9999053832(797) & $-$77202(8) & $-$77217(2) & 15(8) & 116\\
$^{93}$Pd  & $^{93}$Rh & 1.0001147006(758) & $-$59083(8) & $-$58982(370) & $-$101(370) & 132\\
$^{94}$Rh  & $^{94}$Pd & 0.9999219043(1121) & $-$72935(11) & $-$72908(3) & $-$27(12) & 270\\
$^{96}$Ru  & $^{96}$Pd & 0.9998893454(536) & $-$86070(7) & $-$86080(0.2) & 10(6.7) & 122\\
$^{96}$Rh  & $^{96}$Pd & 0.9999608645(160) & $-$79680(5) & $-$79688(10) & 8(11) & 1933\\
$^{96}$Ag  & $^{96}$Pd & 1.0001291921(435) & $-$64641(7) & $-$64512(90) & $-$129(90) & 4171\\
$^{96}$Cd  & $^{96}$Pd & 1.0002310496(1790) & $-$55540(17) & $-$55572(410)$\#$ & 32(410)$\#$ & 8\\
$^{97}$Ru  & $^{97}$Pd & 0.9999080325(2896) & $-$86108(27) & $-$86121(3) & 13(27) & 8\\
$^{97}$Rh  & $^{97}$Pd & 0.9999473290(342) & $-$82561(7) & $-$82598(35) & 37(36) & 539\\
$^{97}$Ag  & $^{97}$Pd & 1.0000761187(74) & $-$70934(6) & $-$70904(12) & $-$30(13) & 89057\\
$^{97}$Cd  & $^{97}$Pd & 1.0001915988(1448) & $-$60509(14) & $-$60734(420) & 225(420) & 117\\
$^{99}$Rh  & $^{99}$Cd & 0.9998309001(2510) & $-$85513(23) & $-$85585(19) & 72(30) & 8\\
$^{99}$Pd  & $^{99}$Cd & 0.9998670610(1187) & $-$82181(11) & $-$82183(5) & 2(12) & 59\\
$^{99}$Ag  & $^{99}$Cd & 0.9999264604(424) & $-$76708(5) & $-$76712(6) & 4(8) & 797\\
$^{99}$In  & $^{99}$Cd & 1.0000923834(825) & $-$61418(9) & $-$61376(298) & $-$42(298) & 118\\
\end{tabular}
\end{ruledtabular}
\end{table*}

Precise mass values were obtained for proton-rich nuclei spanning the $A$\,=\,89–99 isobaric chains, substantially extending the experimental coverage of high-accuracy in the region near the doubly magic nucleus $^{100}$Sn. Figure~\ref{MES} shows the representative mass spectra for $A$\,=\,91, 92, and 96 isobaric series, for which a mass resolving power of $R_m$ $\approx$ 6.0 $\times$ 10$^5$ was achieved. In practice, to avoid misidentification of low-yield ions from contaminants, the analyte ions were measured with different laps in the MRTOF analyzer during the experiment, resulting in distinct TOF distributions that precluded a direct combination of events. To overcome this limitation, the TOF data were converted into a unified mass spectrum, in which the different TOF branches are projected onto a common mass axis, allowing all the detected events to be combined. This procedure enables the clear identification of low-yield isotopes such as $^{92}$Pd and $^{96}$Cd, whose peaks become clearly resolved only after this transformation. To ensure unambiguous identification, a systematic analysis of possible molecular contaminants was performed. No singly charged molecular ions with masses close to those of $^{92}$Pd and $^{96}$Cd were found in this mass region.\par

The mass excess\,(ME) values determined in this work are summarized in Table~\ref{tabME}, with all reference masses taken from AME2020. We report the first precise determinations of ME values of $^{91}$Rh, $^{92}$Pd, and $^{96}$Cd, achieving uncertainties on the order of 10\,keV. For several additional nuclei, including $^{93}$Pd, $^{96}$Ag, $^{97}$Cd, $^{99}$Rh, and $^{99}$In, our measured mass excess values show deviations from the AME2020 evaluations.
The AME2020 values of $^{93}$Pd and $^{97}$Cd were derived indirectly from $\beta$-endpoint measurements\,\cite{Pa19}; our direct measurements therefore provide essential new constraints. The ME value of $^{93}$Pd was recently measured by the FRS group at $-$59127\,(35)\,keV\,\cite{Kr25}  and is consistent with our result. In the case of $^{99}$In, our result is fully consistent with the recent ISOLTRAP value of $-$61429(77)\,keV\,\cite{Mo21}, while reducing the mass uncertainty to 9\,keV. Our ME value for $^{96}$Ag, $-$64641(7)\,keV, is consistent within 3$\sigma$ with the recent high-precision JYFLTRAP Penning trap measurement of $-$64656.69(95)\,keV\,\cite{Ge24}. The deviation observed for $^{99}$Rh exceeds one standard deviation from AME2020, but remains in agreement within 3$\sigma$. We note that the AME2020 value for $^{99}$Rh was derived indirectly from $\beta$-decay measurement\cite{PH69}. The results for the remaining ion species are consistent with the AME2020 evalutions.\par

To evaluate the impact of the newly measured masses on the $rp$-process, we performed one-zone XRB simulations\cite{ko99,Do20}. The calculations follow the thermodynamic evolution and isotopic abundances in a single zone at constant pressure $P=P_{\rm ign}$= 10$^{23.03}$ dyn/cm$^{2}$ , neglecting temperature, density and composition gradients. 
The model parameters and setup are the same as Ref.\,\cite{zhang25}, where $rp$-process reaches the regions with SnSbTe cycle. The reaction network incorporated the newly measured masses of $^{91}$Rh, $^{92}$Pd, and $^{96}$Cd, together with the substantially improved mass values for $^{93}$Pd, $^{97}$Cd, $^{96}$Ag, $^{99}$Rh and $^{99}$In. All relevant proton-capture and inverse reaction rates were recalculated using the updated masses. By varying each mass within its 1$\sigma$ uncertainty, we quantified how the updated mass surface modifies the $rp$-process reaction path and alters the final isotopic composition of the burst ashes.\par

\begin{figure}[ht]
\centering
\vspace{-1.2mm}\includegraphics[width=0.475\textwidth]{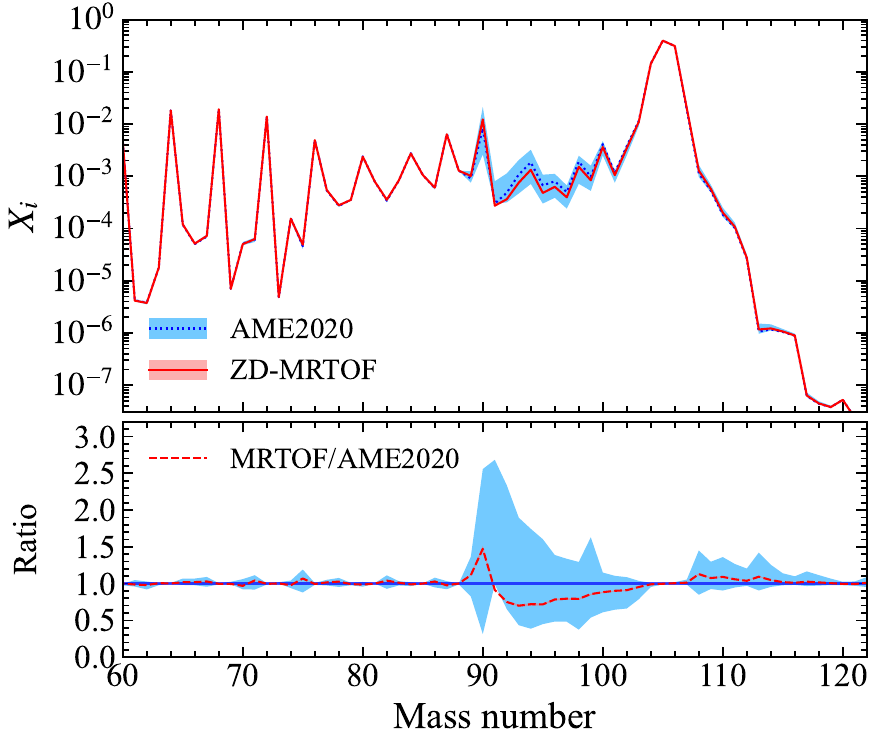}
   \caption{Comparison of the final abundance in ashes from AME2020\,(blue line with shadow) and new masses\,(with 1$\sigma$ uncertainty) measured by the MRTOF\,(red line with shadow). The bottom panel shows the ratio of these final abundances. }
\label{ABU}
\end{figure}

\begin{figure}[ht]
\centering
\includegraphics[width=0.47\textwidth]{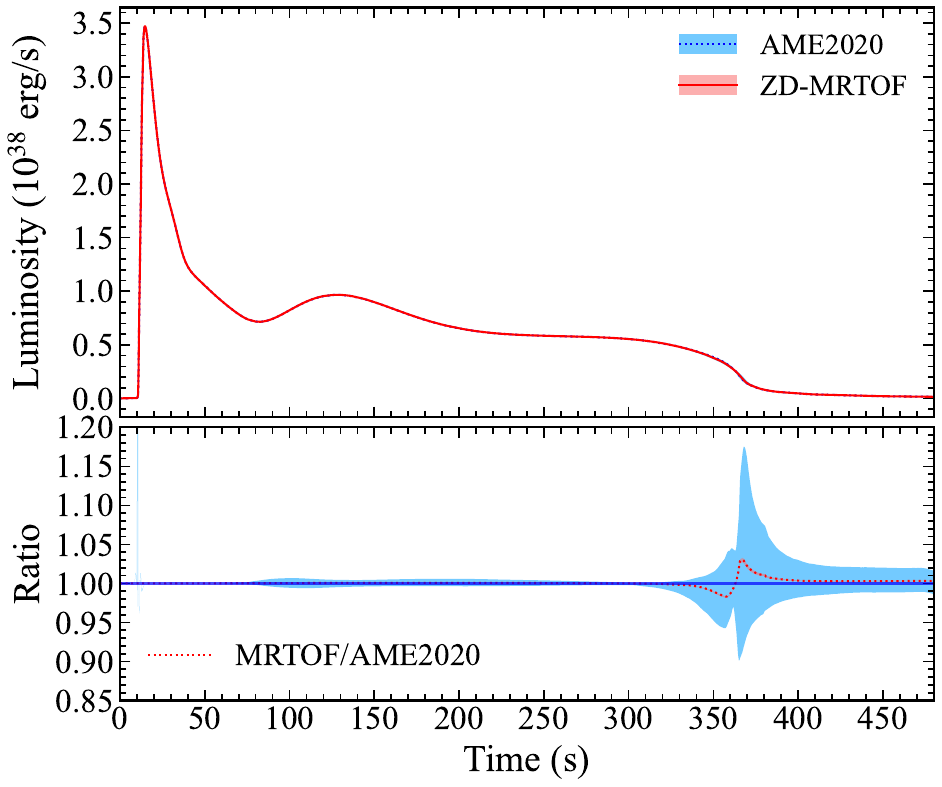}
   \caption{X-ray luminosity as a function of time using AME2020 masses (blue line with shadow) and new masses (red line with shadow, 1$\sigma$ uncertainty). The bottom panel shows the ratio of luminosity. }
\label{LUM}
\end{figure}

Figure~\ref{ABU} shows the final isotopic abundances in the burst ashes. The dashed and solid curves correspond to the calculations using AME2020 mass values and the newly measured masses, respectively, while the red and blue shaded bands indicate the 1$\sigma$ abundance variations originating from the corresponding mass uncertainties. Using AME2020 mass values, the predicted abundances exhibit significant uncertainties of nearly an order of magnitude around $A$\,=\,90 and factors of about five for $90<A<100$. These large uncertainties are dominated by the poorly known mass of  $^{91}$Rh\,(uncertainty:\,298\,keV)\,\cite{Sc17}. Because the $^{90}$Ru$(p,\gamma)$$^{91}$Rh reaction has a small $Q$ value, a 300\,keV mass uncertainty translates into a variation of the reverse photodisintegration rate by about a factor of 30 at $T$\,=\,1\,GK, which is the corresponding temperature in our trajectories. Our new mass measurement drastically reduces this uncertainty. Moreover, the newly determined mass of $^{91}$Rh yields a smaller $Q$ value of 0.86\,MeV, compared to 0.98\,MeV from AME2020, which enhances the $\beta$$^{+}$ decay flow of $^{90}$Ru relative to proton capture. This shift amplifies the final abundance at $A=90$ and suppresses the production of heavier nuclei\,(Fig.~\ref{Path}\,(a)). In addition to $^{91}$Rh, the mass of $^{97}$Cd has been demonstrated to play an important role on XRBs studies since their $(p, \gamma)$ reactions also have $Q\leq1\rm \,MeV$\,\cite{Pa09,Pa13}. 
With AME2020 masses, the resulting uncertainties in the final abundances of the $A$\,=\,98,\,99 nuclei reach about 30\%. Our new measurements reduce the mass uncertainties of these nuclei to the order of 10\,keV, lowering the corresponding abundance uncertainties. We compare the calculated XRB light curves as shown in Figure ~\ref{LUM}. The new mass values significantly reduce the uncertainty of the late-time light-curve tail, in particular around $t$\,$\approx$\,350\,s.

We also investigated the impact of the new mass measurements on the $\nu$$p$-process by modeling neutrino-driven winds using a general-relativistic, steady-state, and spherically symmetric framework\,\cite{qi96, ot00}. In our calculation, the mass and radius of the proto-neutron star, the temperature and luminosity of the neutrino-driven wind are consistent with 3D core-collapse supernova simulations \cite{bu20}. To explore the sensitivity to different astrophysical conditions, we considered a range of progenitors and initial setups, covering electron fractions
$0.5 \leq Y_e \leq 0.7$. The influence of the new masses was quantified by incorporating the corresponding changes in reaction $Q$ values into the reaction network and evaluating their impact on the predicted isotopic abundances.\par
The $\nu$$p$-process proceeds in proton-rich regions near the line of stability. Among the nuclei measured in this work, $^{99}$Rh lies on the $\nu$$p$-process reaction path and influences the final abundances. Our first direct mass determination of $^{99}$Rh is 72\,keV higher than the value determined by the $\beta$-endpoint measurement\cite{PH69}, increasing the $^{99}$Rh($n,\gamma$)$^{100}$Rh reaction rate by 2.7\%. Model calculations show that only heavier proto-neutron stars ($\approx 2.0\,M_{\odot}$) produce significant $^{99}$Rh yields, with a $^{99}$Rh/H ratio exceeding 10$^{-10}$. The enhanced reaction rate increases the final abundances of $^{102}$Pd and $^{104}$Pd by about 1.5\%. We find that, within the mass region covered by this work, $^{99}$Rh plays a significant role in the $\nu$$p$-process reaction flow. 
It should be noted, however, that the detailed reaction flow depends sensitively on the underlying astrophysical conditions. For example,  Wanajo \textit{et al.}\ \cite{Wa11} showed that adopting a wind-termination radius of 300 km shifts the $\nu p$-process path further from stability, potentially redirecting the flow through more proton-rich nuclei such as $^{91}$Rh, $^{92}$Pd, and $^{93}$Pd. A systematic investigation of such environmental dependencies will be presented in a forthcoming study.

In summary, we have performed high-precision mass measurements of proton-rich nuclei near the doubly magic nucleus $^{100}$Sn, substantially extending experimental coverage along the $A$ = 89-99 isobaric chains. We report the first precise mass determinations of $^{91}$Rh, $^{92}$Pd, and $^{96}$Cd, together with the significantly improved mass values for several additional nuclei of astrophysical relevance. Incorporating the new masses into XRB simulations, we demonstrate that the previously large uncertainties in abundances around $A$\,=\,90, dominated by the poorly known mass of $^{91}$Rh, are significantly reduced. The reaction flow is consequently shifted toward enhanced production at $A$\,=\,90 with suppressed synthesis of heavier nuclei. We further show that, within the mass region studied in this work, $^{99}$Rh plays a significant role in the $\nu$$p$-process reaction flow.
This highlights the necessity for future high-precision mass measurements closer to the line of stability to achieve a more comprehensive understanding of the abundances of light $p$-nuclei. Beyond their astrophysical relevance, the masses measured in this work also provide valuable benchmarks for investigating shell evolution near 
$^{100}$Sn; a detailed discussion will be presented in a forthcoming publication.

\begin{acknowledgments}
	We are grateful to the RIKEN Nishina Center for Accelerator-based Science and the Center for Nuclear Study at the University of Tokyo. This work was supported by Research Grants Council (RGC) of Hong Kong (GRF-17300625, GRF-17312522) and the National Natural Science Foundation of China (Grant Number:12525507 and 12135004), the Japan Society for the Promotion of Science, KAKENHI, Grant Numbers 17H06090, 20H05648, 22H01257, 22H04946, 25H01273,19K14750, and 21K13951, and also RIKEN programs (r-EMU and RiNA-Net). Y.L. is supported by Boya fellowship of Peking University and the China Postdoctoral Science Foundation under Grant Number 2025T180924. A.D. is supported by JSPS KAKENHI (Grant No. J25K17403). T.K. is supported by the National Key R\&D Program of China (2022YFA1602401) and the National Natural Science Foundation of China (No. 12335009 and 12435010).  
\end{acknowledgments}

\bibliography{refs}

@article{Be16,
title = {Frontiers in nuclear astrophysics},
journal = {Progress in Particle and Nuclear Physics},
volume = {89},
pages = {56-100},
year = {2016},
issn = {0146-6410},
doi = {https://doi.org/10.1016/j.ppnp.2016.04.001},
url = {https://www.sciencedirect.com/science/article/pii/S0146641016300011},
author = {C.A. Bertulani and T. Kajino}
}

@article{Wo76,
  title={$\gamma$-ray bursts from thermonuclear explosions on neutron stars},
  author={Woosley, SE and Taam, Ronald E},
  journal={Nature},
  volume={263},
  number={5573},
  pages={101--103},
  year={1976},
  doi = {https://doi.org/10.1038/263101a0},
  publisher={Nature Publishing Group UK London}
}

@article{Jo77,
  title={X-ray bursts and neutron-star thermonuclear flashes},
  author={Joss, PC},
  journal={Nature},
  volume={270},
  number={5635},
  pages={310--314},
  year={1977},
  doi = {https://doi.org/10.1038/270310a0},
  publisher={Nature Publishing Group UK London}
}

@article{Pa13,
title = {Nucleosynthesis in type I X-ray bursts},
journal = {Progress in Particle and Nuclear Physics},
volume = {69},
pages = {225-253},
year = {2013},
issn = {0146-6410},
doi = {https://doi.org/10.1016/j.ppnp.2012.11.002},
url = {https://www.sciencedirect.com/science/article/pii/S0146641012001354},
author = {A. Parikh and J. José and G. Sala and C. Iliadis}
}

@article{Me18,
doi = {10.1088/1361-6471/aad171},
url = {https://doi.org/10.1088/1361-6471/aad171},
year = {2018},
month = {jul},
publisher = {IOP Publishing},
volume = {45},
number = {9},
pages = {093001},
author = {Meisel, Zach and Deibel, Alex and Keek, Laurens and Shternin, Peter and Elfritz, Justin},
title = {Nuclear physics of the outer layers of accreting neutron stars},
journal = {Journal of Physics G: Nuclear and Particle Physics}
}

@article{Sc98,
title = {rp-process nucleosynthesis at extreme temperature and density conditions},
journal = {Physics Reports},
volume = {294},
number = {4},
pages = {167-263},
year = {1998},
issn = {0370-1573},
doi = {https://doi.org/10.1016/S0370-1573(97)00048-3},
url = {https://www.sciencedirect.com/science/article/pii/S0370157397000483},
author = {H. Schatz and A. Aprahamian and J. Görres and M. Wiescher and T. Rauscher and J.F. Rembges and F.-K. Thielemann and B. Pfeiffer and P. Möller and K.-L. Kratz and H. Herndl and B.A. Brown and H. Rebel}
}

@article{Wa81,
   author = {Wallace, R. K. and Woosley, S. E.},
   title = {Explosive hydrogen burning},
   journal = {The Astrophysical Journal Supplement Series},
   volume = {45},
   pages = {389-420},
   ISSN = {0067-0049},
   DOI = {10.1086/190717},
   url = {https://ui.adsabs.harvard.edu/abs/1981ApJS...45..389W},
   year = {1981},
   type = {Journal Article}
}

@ARTICLE{Wo94,
       author = {{van Wormer}, L. and {G{\"o}rres}, J. and {Iliadis}, C. and {Wiescher}, M. and {Thielemann}, F. -K.},
        title = "{Reaction Rates and Reaction Sequences in the rp-Process}",
      journal = {\apj},
         year = 1994,
        month = sep,
       volume = {432},
        pages = {326},
          doi = {10.1086/174572},
       adsurl = {https://ui.adsabs.harvard.edu/abs/1994ApJ...432..326V}
}

@article{Wo04,
doi = {10.1086/381533},
url = {https://dx.doi.org/10.1086/381533},
year = {2004},
month = {mar},
publisher = {},
volume = {151},
number = {1},
pages = {75},
author = {S. E. Woosley and A. Heger and A. Cumming and R. D. Hoffman and J. Pruet and T. Rauscher and J. L. Fisker and H. Schatz and B. A. Brown and M. Wiescher},
title = {Models for Type I X-Ray Bursts with Improved Nuclear Physics},
journal = {The Astrophysical Journal Supplement Series},
}

@article{Fi08,
doi = {10.1086/521104},
url = {https://dx.doi.org/10.1086/521104},
year = {2008},
month = {jan},
publisher = {},
volume = {174},
number = {1},
pages = {261},
author = {Jacob Lund Fisker and Hendrik Schatz and Friedrich-Karl Thielemann},
title = {Explosive Hydrogen Burning during Type I X-Ray Bursts},
journal = {The Astrophysical Journal Supplement Series}
}

@article{Sc01,
  title = {End Point of the $\mathit{rp}$ Process on Accreting Neutron Stars},
  author = {Schatz, H. and Aprahamian, A. and Barnard, V. and Bildsten, L. and Cumming, A. and Ouellette, M. and Rauscher, T. and Thielemann, F.-K. and Wiescher, M.},
  journal = {Phys. Rev. Lett.},
  volume = {86},
  issue = {16},
  pages = {3471--3474},
  numpages = {0},
  year = {2001},
  month = {Apr},
  publisher = {American Physical Society},
  doi = {10.1103/PhysRevLett.86.3471},
  url = {https://link.aps.org/doi/10.1103/PhysRevLett.86.3471}
}

@article{We06,
doi = {10.1086/499426},
url = {https://doi.org/10.1086/499426},
year = {2006},
month = {mar},
publisher = {},
volume = {639},
number = {2},
pages = {1018},
author = {Weinberg, Nevin N. and Bildsten, Lars and Schatz, Hendrik},
title = {Exposing the Nuclear Burning Ashes of Radius Expansion Type I X-Ray Bursts},
journal = {The Astrophysical Journal}
}

@article{Wa06,
doi = {10.1086/505483},
url = {https://doi.org/10.1086/505483},
year = {2006},
month = {aug},
publisher = {},
volume = {647},
number = {2},
pages = {1323},
author = {Wanajo, Shinya},
title = {The rp-Process in Neutrino-driven Winds},
journal = {The Astrophysical Journal}
}

@article{Fr06,
  title = {Neutrino-Induced Nucleosynthesis of {A} $>$ 64 Nuclei: The $\ensuremath{\nu}p$ Process},
  author = {Fr\"ohlich, C. and Mart\'{\i}nez-Pinedo, G. and Liebend\"orfer, M. and Thielemann, F.-K. and Bravo, E. and Hix, W. R. and Langanke, K. and Zinner, N. T.},
  journal = {Phys. Rev. Lett.},
  volume = {96},
  issue = {14},
  pages = {142502},
  numpages = {4},
  year = {2006},
  month = {Apr},
  publisher = {American Physical Society},
  doi = {10.1103/PhysRevLett.96.142502},
  url = {https://link.aps.org/doi/10.1103/PhysRevLett.96.142502}
}

@article{Pr06,
doi = {10.1086/503891},
url = {https://doi.org/10.1086/503891},
year = {2006},
month = {jun},
publisher = {},
volume = {644},
number = {2},
pages = {1028},
author = {Pruet, J. and Hoffman, R. D. and Woosley, S. E. and Janka, H.-T. and Buras, R.},
title = {Nucleosynthesis in Early Supernova Winds. II. The Role of Neutrinos},
journal = {The Astrophysical Journal}
}

@article{ko04,
  title={Final products of the rp-process on accreting neutron stars},
  author={Koike, Osamu and Hashimoto, Masa-aki and Kuromizu, Reiko and Fujimoto, Shin-ichirou},
  journal={The Astrophysical Journal},
  volume={603},
  number={1},
  pages={242},
  year={2004},
  doi = {10.1086/381354 },
  publisher={IOP Publishing}
}

@article{Ar03,
title = {The p-process of stellar nucleosynthesis: astrophysics and nuclear physics status},
journal = {Physics Reports},
volume = {384},
number = {1},
pages = {1-84},
year = {2003},
issn = {0370-1573},
doi = {https://doi.org/10.1016/S0370-1573(03)00242-4},
url = {https://www.sciencedirect.com/science/article/pii/S0370157303002424},
author = {M. Arnould and S. Goriely}
}

@article{Sc99,
doi = {10.1086/307837},
url = {https://doi.org/10.1086/307837},
year = {1999},
month = {oct},
publisher = {},
volume = {524},
number = {2},
pages = {1014},
author = {Schatz, Hendrik and Bildsten, Lars and Cumming, Andrew and Wiescher, Michael},
title = {The Rapid Proton Process Ashes from Stable Nuclear Burning on an Accreting Neutron Star},
journal = {The Astrophysical Journal}
}

@article{Sc17,
doi = {10.3847/1538-4357/aa7de9},
url = {https://doi.org/10.3847/1538-4357/aa7de9},
year = {2017},
month = {aug},
publisher = {The American Astronomical Society},
volume = {844},
number = {2},
pages = {139},
author = {Schatz, H. and Ong, W.-J.},
title = {Dependence of X-Ray Burst Models on Nuclear Masses},
journal = {The Astrophysical Journal}
}

@inproceedings{Fr12,
  title={Reaction rate uncertainties and the $\nu$ p-process},
  author={Fr{\"o}hlich, C and Rauscher, T},
  booktitle={AIP Conference Proceedings},
  volume={1484},
  number={1},
  pages={232--239},
  year={2012},
  doi = {https://doi.org/10.1063/1.4763400},
  organization={American Institute of Physics}
}

@article{Ya07,
title = {The RIKEN RI Beam Factory Project: A status report},
journal = {Nuclear Instruments and Methods in Physics Research Section B: Beam Interactions with Materials and Atoms},
volume = {261},
number = {1},
pages = {1009-1013},
year = {2007},
note = {The Application of Accelerators in Research and Industry},
issn = {0168-583X},
doi = {https://doi.org/10.1016/j.nimb.2007.04.174},
url = {https://www.sciencedirect.com/science/article/pii/S0168583X07009792},
author = {Yasushige Yano}
}

@article{Ku14,
title = {In-flight RI beam separator BigRIPS at RIKEN and elsewhere in Japan},
journal = {Nuclear Instruments and Methods in Physics Research Section B: Beam Interactions with Materials and Atoms},
volume = {204},
pages = {97-113},
year = {2003},
issn = {0168-583X},
doi = {https://doi.org/10.1016/S0168-583X(02)01896-7},
url = {https://www.sciencedirect.com/science/article/pii/S0168583X02018967},
author = {Toshiyuki Kubo}
}

@article{Ku12,
    author = {Kubo, Toshiyuki and Kameda, Daisuke and Suzuki, Hiroshi and Fukuda, Naoki and Takeda, Hiroyuki and Yanagisawa, Yoshiyuki and Ohtake, Masao and Kusaka, Kensuke and Yoshida, Koichi and Inabe, Naohito and Ohnishi, Tetsuya and Yoshida, Atsushi and Tanaka, Kanenobu and Mizoi, Yutaka},
    title = {BigRIPS separator and ZeroDegree spectrometer at RIKEN RI Beam Factory},
    journal = {Progress of Theoretical and Experimental Physics},
    volume = {2012},
    number = {1},
    pages = {03C003},
    year = {2012},
    month = {12},
    issn = {2050-3911},
    doi = {10.1093/ptep/pts064},
    url = {https://doi.org/10.1093/ptep/pts064},
}

@article{Ro24,
title = {The new MRTOF mass spectrograph following the ZeroDegree spectrometer at RIKEN’s RIBF facility},
journal = {Nuclear Instruments and Methods in Physics Research Section A: Accelerators, Spectrometers, Detectors and Associated Equipment},
volume = {1047},
pages = {167824},
year = {2023},
issn = {0168-9002},
doi = {https://doi.org/10.1016/j.nima.2022.167824},
url = {https://www.sciencedirect.com/science/article/pii/S0168900222011160},
author = {M. Rosenbusch and M. Wada and S. Chen and A. Takamine and S. Iimura and D. Hou and W. Xian and S. Yan and P. Schury and Y. Hirayama and Y. Ito and H. Ishiyama and S. Kimura and T. Kojima and J. Lee and J. Liu and S. Michimasa and H. Miyatake and J.Y. Moon and M. Mukai and S. Naimi and S. Nishimura and T. Niwase and T. Sonoda and Y.X. Watanabe and H. Wollnik}
}

@article{Ho23,
  title = {First direct mass measurement for neutron-rich $^{112}\mathrm{Mo}$ with the new ZD-MRTOF mass spectrograph system},
  author = {Hou, D. S. and Takamine, A. and Rosenbusch, M. and Xian, W. D. and Iimura, S. and Chen, S. D. and Wada, M. and Ishiyama, H. and Schury, P. and Niu, Z. M. and Liang, H. Z. and Yan, S. X. and Doornenbal, P. and Hirayama, Y. and Ito, Y. and Kimura, S. and Kojima, T. M. and Korten, W. and Lee, J. and Liu, J. J. and Liu, Z. and Michimasa, S. and Miyatake, H. and Moon, J. Y. and Naimi, S. and Nishimura, S. and Niwase, T. and Sonoda, T. and Suzuki, D. and Watanabe, Y. X. and Wimmer, K. and Wollnik, H.},
  journal = {Phys. Rev. C},
  volume = {108},
  issue = {5},
  pages = {054312},
  numpages = {10},
  year = {2023},
  month = {Nov},
  publisher = {American Physical Society},
  doi = {10.1103/PhysRevC.108.054312},
  url = {https://link.aps.org/doi/10.1103/PhysRevC.108.054312}
}

@article{It18,
  title = {First Direct Mass Measurements of Nuclides around {Z} = 100 with a Multireflection Time-of-Flight Mass Spectrograph},
  author = {Ito, Y. and Schury, P. and Wada, M. and Arai, F. and Haba, H. and Hirayama, Y. and Ishizawa, S. and Kaji, D. and Kimura, S. and Koura, H. and MacCormick, M. and Miyatake, H. and Moon, J. Y. and Morimoto, K. and Morita, K. and Mukai, M. and Murray, I. and Niwase, T. and Okada, K. and Ozawa, A. and Rosenbusch, M. and Takamine, A. and Tanaka, T. and Watanabe, Y. X. and Wollnik, H. and Yamaki, S.},
  journal = {Phys. Rev. Lett.},
  volume = {120},
  issue = {15},
  pages = {152501},
  numpages = {6},
  year = {2018},
  month = {Apr},
  publisher = {American Physical Society},
  doi = {10.1103/PhysRevLett.120.152501},
  url = {https://link.aps.org/doi/10.1103/PhysRevLett.120.152501}
}

@article{Wa03,
title = {Slow RI-beams from projectile fragment separators},
journal = {Nuclear Instruments and Methods in Physics Research Section B: Beam Interactions with Materials and Atoms},
volume = {204},
pages = {570-581},
year = {2003},
issn = {0168-583X},
doi = {https://doi.org/10.1016/S0168-583X(02)02151-1},
url = {https://www.sciencedirect.com/science/article/pii/S0168583X02021511},
author = {Michiharu Wada and Yoshihisa Ishida and Takashi Nakamura and Yasunori Yamazaki and Tadashi Kambara and Hitoshi Ohyama and Yasushi Kanai and Takao M. Kojima and Youichi Nakai and Nagayasu Ohshima and Atsushi Yoshida and Toshiyuki Kubo and Yukari Matsuo and Yoshimitsu Fukuyama and Kunihiro Okada and Tetsu Sonoda and Shunsuke Ohtani and Koji Noda and Hirokane Kawakami and Ichiro Katayama}
}

@article{Ta05,
    author = {Takamine, A. and Wada, M. and Ishida, Y. and Nakamura, T. and Okada, K. and Yamazaki, Y. and Kambara, T. and Kanai, Y. and Kojima, T. M. and Nakai, Y. and Oshima, N. and Yoshida, A. and Kubo, T. and Ohtani, S. and Noda, K. and Katayama, I. and Hostain, P. and Varentsov, V. and Wollnik, H.},
    title = {Space-charge effects in the catcher gas cell of a rf ion guide},
    journal = {Review of Scientific Instruments},
    volume = {76},
    number = {10},
    pages = {103503},
    year = {2005},
    month = {10},
    issn = {0034-6748},
    doi = {10.1063/1.2090290},
    url = {https://doi.org/10.1063/1.2090290}
}

@article{Bo11,
title = {``Ion surfing'' with radiofrequency carpets},
journal = {International Journal of Mass Spectrometry},
volume = {299},
number = {2},
pages = {131-138},
year = {2011},
issn = {1387-3806},
doi = {https://doi.org/10.1016/j.ijms.2010.09.032},
url = {https://www.sciencedirect.com/science/article/pii/S1387380610003659},
author = {G. Bollen}
}

@article{Ar14,
title = {Investigation of the ion surfing transport method with a circular rf carpet},
journal = {International Journal of Mass Spectrometry},
volume = {362},
pages = {56-58},
year = {2014},
issn = {1387-3806},
doi = {https://doi.org/10.1016/j.ijms.2014.01.005},
url = {https://www.sciencedirect.com/science/article/pii/S1387380614000098},
author = {F. Arai and Y. Ito and M. Wada and P. Schury and T. Sonoda and H. Mita}
}

@article{Xi25,
  title={Atomic mass measurements of neutron-rich nuclides on the path to 78 Ni with a $\beta$-TOF-equipped MRTOF},
  author={Xian, W and Rosenbusch, M and Phong, VH and Wada, M and Schury, P and Hou, D and Takamine, A and Chen, S and Niwase, T and Hirayama, Y and others},
  journal={Frontiers in Physics},
  volume={13},
  pages={1644477},
  year={2025},
  doi={ https://doi.org/10.3389/fphy.2025.1644477},
  publisher={Frontiers}
}

@article{Ni23,
author = {Niwase, Toshitaka and Xian, Wenduo and Wada, Michiharu and Rosenbusch, Marco and Chen, Sidong and Takamine, Aiko and Liu, Jiajian and Iimura, Shun and Hou, Dongsheng and Yan, Shuxiong and Ishiyama, Hironobu and Miyatake, Hiroari and Nishimura, Shunji and Kaji, Daiya and Morimoto, Kouji and Hirayama, Yoshikazu and Watanabe, Yutaka X and Kimura, Sota and Schury, Peter and Wollnik, Hermann},
title = {Development of a $\beta$-TOF detector: An enhancement of the $\alpha$-TOF detector for use with $\beta$-decaying nuclides},
journal = {Progress of Theoretical and Experimental Physics},
volume = {2023},
number = {3},
pages = {031H01},
year = {2023},
month = {03},
issn = {2050-3911},
doi = {10.1093/ptep/ptad039},
url = {https://doi.org/10.1093/ptep/ptad039}
}

@article{Sc14,
title = {A high-resolution multi-reflection time-of-flight mass spectrograph for precision mass measurements at RIKEN/SLOWRI},
journal = {Nuclear Instruments and Methods in Physics Research Section B: Beam Interactions with Materials and Atoms},
volume = {335},
pages = {39-53},
year = {2014},
issn = {0168-583X},
doi = {https://doi.org/10.1016/j.nimb.2014.05.016},
url = {https://www.sciencedirect.com/science/article/pii/S0168583X1400559X},
author = {P. Schury and M. Wada and Y. Ito and F. Arai and S. Naimi and T. Sonoda and H. Wollnik and V.A. Shchepunov and C. Smorra and C. Yuan}
}

@article{Ro18,
  title = {New mass anchor points for neutron-deficient heavy nuclei from direct mass measurements of radium and actinium isotopes},
  author = {Rosenbusch, M. and Ito, Y. and Schury, P. and Wada, M. and Kaji, D. and Morimoto, K. and Haba, H. and Kimura, S. and Koura, H. and MacCormick, M. and Miyatake, H. and Moon, J. Y. and Morita, K. and Murray, I. and Niwase, T. and Ozawa, A. and Reponen, M. and Takamine, A. and Tanaka, T. and Wollnik, H.},
  journal = {Phys. Rev. C},
  volume = {97},
  issue = {6},
  pages = {064306},
  numpages = {8},
  year = {2018},
  month = {Jun},
  publisher = {American Physical Society},
  doi = {10.1103/PhysRevC.97.064306},
  url = {https://link.aps.org/doi/10.1103/PhysRevC.97.064306}
}

@article{Wa21,
doi = {10.1088/1674-1137/abddaf},
url = {https://doi.org/10.1088/1674-1137/abddaf},
year = {2021},
month = {mar},
publisher = {Chinese Physical Society and the Institute of High Energy Physics of the Chinese Academy of Sciences and the Institute of Modern Physics of the Chinese Academy of Sciences and IOP Publishing Ltd},
volume = {45},
number = {3},
pages = {030003},
author = {Wang, Meng and Huang, W.J. and Kondev, F.G. and Audi, G. and Naimi, S.},
title = {The AME 2020 atomic mass evaluation (II). Tables, graphs and references*},
journal = {Chinese Physics C}
}

@article{Pa19,
  title = {New and comprehensive $\ensuremath{\beta}$- and $\ensuremath{\beta}p$-decay spectroscopy results in the vicinity of $^{100}\mathrm{Sn}$},
  author = {Park, J. and Kr\"ucken, R. and Lubos, D. and Gernh\"auser, R. and Lewitowicz, M. and Nishimura, S. and Ahn, D. S. and Baba, H. and Blank, B. and Blazhev, A. and Boutachkov, P. and Browne, F. and \ifmmode \check{C}\else \v{C}\fi{}elikovi\ifmmode \acute{c}\else \'{c}\fi{}, I. and de France, G. and Doornenbal, P. and Faestermann, T. and Fang, Y. and Fukuda, N. and Giovinazzo, J. and Goel, N. and G\'orska, M. and Grawe, H. and Ilieva, S. and Inabe, N. and Isobe, T. and Jungclaus, A. and Kameda, D. and Kim, G. D. and Kim, Y.-K. and Kojouharov, I. and Kubo, T. and Kurz, N. and Kwon, Y. K. and Lorusso, G. and Moschner, K. and Murai, D. and Nishizuka, I. and Patel, Z. and Rajabali, M. M. and Rice, S. and Sakurai, H. and Schaffner, H. and Shimizu, Y. and Sinclair, L. and S\"oderstr\"om, P.-A. and Steiger, K. and Sumikama, T. and Suzuki, H. and Takeda, H. and Wang, Z. and Watanabe, H. and Wu, J. and Xu, Z. Y.},
  journal = {Phys. Rev. C},
  volume = {99},
  issue = {3},
  pages = {034313},
  numpages = {26},
  year = {2019},
  month = {Mar},
  publisher = {American Physical Society},
  doi = {10.1103/PhysRevC.99.034313},
  url = {https://link.aps.org/doi/10.1103/PhysRevC.99.034313}
}

@article{Mo21,
  title={Mass measurements of 99--101In challenge ab initio nuclear theory of the nuclide 100Sn},
  author={Mougeot, M and Atanasov, D and Karthein, J and Wolf, RN and Ascher, P and Blaum, K and Chrysalidis, K and Hagen, Gaute and Holt, JD and Huang, WJ and others},
  journal={Nature Physics},
  volume={17},
  number={10},
  pages={1099--1103},
  year={2021},
  doi={https://doi.org/10.1038/s41567-021-01326-9},
  publisher={Nature Publishing Group UK London}
}

@article{Ge24,
  title = {High-Precision Mass Measurements of Neutron Deficient Silver Isotopes Probe the Robustness of the $N=50$ Shell Closure},
  author = {Ge, Zhuang and Reponen, Mikael and Eronen, Tommi and Hu, Baishan and Kortelainen, Markus and Kankainen, Anu and Moore, Iain and Nesterenko, Dmitrii and Yuan, Cenxi and Beliuskina, Olga and Ca\~nete, Laetitia and de Groote, Ruben and Delafosse, Cl\'ement and Dickel, Timo and de Roubin, Antoine and Geldhof, Sarina and Gins, Wouter and Holt, Jason D. and Hukkanen, Marjut and Jaries, Arthur and Jokinen, Ari and Koszor\'us, \'Agota and Kripk\'o-Koncz, Gabriella and Kujanp\"a\"a, Sonja and Lam, Yi Hua and Nikas, Stylianos and Ortiz-Cortes, Alejandro and Penttil\"a, Heikki and Pitman-Weymouth, Daniel and Pla\ss{}, Wolfgang and Pohjalainen, Ilkka and Raggio, Andrea and Rinta-Antila, Sami and Romero, Jorge and Stryjczyk, Marek and Vilen, Markus and Virtanen, Ville and Zadvornaya, Alexandra},
  journal = {Phys. Rev. Lett.},
  volume = {133},
  issue = {13},
  pages = {132503},
  numpages = {9},
  year = {2024},
  month = {Sep},
  publisher = {American Physical Society},
  doi = {10.1103/PhysRevLett.133.132503},
  url = {https://link.aps.org/doi/10.1103/PhysRevLett.133.132503}
}

@article{Do20,
    author = {Dohi, Akira and Hashimoto, Masa-aki and Yamada, Rio and Matsuo, Yasuhide and Fujimoto, Masayuki Y},
    title = {An approach to constrain models of accreting neutron stars with the use of an equation of state},
    journal = {Progress of Theoretical and Experimental Physics},
    volume = {2020},
    number = {3},
    pages = {033E02},
    year = {2020},
    month = {03},
    issn = {2050-3911},
    doi = {10.1093/ptep/ptaa010},
    url = {https://doi.org/10.1093/ptep/ptaa010}
}

@article{Xi23,
  title = {Isochronous mass measurements of neutron-deficient nuclei from $^{112}\mathrm{Sn}$ projectile fragmentation},
  author = {Xing, Y. M. and Yuan, C. X. and Wang, M. and Zhang, Y. H. and Zhou, X. H. and Litvinov, Yu. A. and Blaum, K. and Xu, H. S. and Bao, T. and Chen, R. J. and Fu, C. Y. and Gao, B. S. and Ge, W. W. and He, J. J. and Huang, W. J. and Liao, T. and Li, J. G. and Li, H. F. and Litvinov, S. and Naimi, S. and Shuai, P. and Sun, M. Z. and Wang, Q. and Xu, X. and Xu, F. R. and Yamaguchi, T. and Yan, X. L. and Yang, J. C. and Yuan, Y. J. and Zeng, Q. and Zhang, M. and Zhou, X.},
  journal = {Phys. Rev. C},
  volume = {107},
  issue = {1},
  pages = {014304},
  numpages = {14},
  year = {2023},
  month = {Jan},
  publisher = {American Physical Society},
  doi = {10.1103/PhysRevC.107.014304},
  url = {https://link.aps.org/doi/10.1103/PhysRevC.107.014304}
}

@article{Ho20,
title = {Isomer studies in the vicinity of the doubly-magic nucleus 100Sn: Observation of a new low-lying isomeric state in 97Ag},
journal = {Physics Letters B},
volume = {802},
pages = {135200},
year = {2020},
issn = {0370-2693},
doi = {https://doi.org/10.1016/j.physletb.2020.135200},
url = {https://www.sciencedirect.com/science/article/pii/S0370269320300046},
author = {Christine Hornung and Daler Amanbayev and Irene Dedes and Gabriella Kripko-Koncz and Ivan Miskun and Noritaka Shimizu and Samuel {Ayet San Andrés} and Julian Bergmann and Timo Dickel and Jerzy Dudek and Jens Ebert and Hans Geissel and Magdalena Górska and Hubert Grawe and Florian Greiner and Emma Haettner and Takaharu Otsuka and Wolfgang R. Plaß and Sivaji Purushothaman and Ann-Kathrin Rink and Christoph Scheidenberger and Helmut Weick and Soumya Bagchi and Andrey Blazhev and Olga Charviakova and Dominique Curien and Andrew Finlay and Satbir Kaur and Wayne Lippert and Jan-Hendrik Otto and Zygmunt Patyk and Stephane Pietri and Yoshiki K. Tanaka and Yusuke Tsunoda and John S. Winfield}
}

@article{ko99,
title = "Rapid proton capture on accreting neutron stars -effects of uncertainty in the nuclear process",
author = "O. Koike and M. Hashimoto and K. Arai and S. Wanajo",
year = "1999",
language = "English",
volume = "342",
pages = "464--473",
journal = "Astronomy and Astrophysics",
issn = "0004-6361",
publisher = "EDP Sciences",
number = "2",
URL={https://adsabs.harvard.edu/full/1999A%26A...342..464K}
}

@article{zhang25,
  title = {Impact of the experimental mass of $^{70}\mathrm{Kr}$ on the $^{68}\mathrm{Se}$ waiting point in the $rp$ process},
  author = {Zhang, M. and Luo, Y. and Dohi, A. and Xu, X. and Yan, X. L. and Kajino, T. and Zhang, Y. H. and Wang, M.},
  journal = {Phys. Rev. C},
  volume = {112},
  issue = {6},
  pages = {065808},
  numpages = {6},
  year = {2025},
  month = {Dec},
  publisher = {American Physical Society},
  doi = {10.1103/kgw8-h6qz},
  url = {https://link.aps.org/doi/10.1103/kgw8-h6qz}
}

@article{Pa09,
  title = {Impact of uncertainties in reaction $Q$ values on nucleosynthesis in type I x-ray bursts},
  author = {Parikh, A. and Jos\'e, J. and Iliadis, C. and Moreno, F. and Rauscher, T.},
  journal = {Phys. Rev. C},
  volume = {79},
  issue = {4},
  pages = {045802},
  numpages = {12},
  year = {2009},
  month = {Apr},
  publisher = {American Physical Society},
  doi = {10.1103/PhysRevC.79.045802},
  url = {https://link.aps.org/doi/10.1103/PhysRevC.79.045802}
}

@article{Pag12,
    author = "Page, Dany and Reddy, Sanjay",
    title = "{Thermal and transport properties of the neutron star inner crust}",
    journal = {},
    eprint = "1201.5602",
    archivePrefix = "arXiv",
    primaryClass = "nucl-th",
    reportNumber = "INT-PUB-12-002",
    month = "1",
    year = "2012"
}

@article{qi96,
  title={Nucleosynthesis in neutrino-driven winds. I. The physical conditions},
  author={Qian, Y-Z and Woosley, SE},
  journal={The Astrophysical Journal},
  volume={471},
  number={1},
  pages={331},
  year={1996},
  publisher={IOP Publishing},
DOI = {10.1086/177973}
}

@article{ot00,
  title={General relativistic effects on neutrino-driven winds from young, hot neutron stars and r-process nucleosynthesis},
  author={Otsuki, Kaori and Tagoshi, Hideyuki and Kajino, Toshitaka and Wanajo, Shin-ya},
  journal={The Astrophysical Journal},
  volume={533},
  number={1},
  pages={424},
  year={2000},
  publisher={IOP Publishing},
DOI = {10.1086/308632}
}

@article{bu20,
  title={{The overarching framework of core-collapse supernova explosions as revealed by 3D FORNAX simulations}},
  author={Burrows, Adam and Radice, David and Vartanyan, David and Nagakura, Hiroki and Skinner, M Aaron and Dolence, Joshua C},
  author={Burrows, Adam and Radice, David and Vartanyan, David and Nagakura, Hiroki and Skinner, M Aaron and Dolence, Joshua C},
  all_journal={Monthly Notices of the Royal Astronomical Society},
  journal={MNRAS},
  volume={491},
  number={2},
  pages={2715--2735},
  year={2020},
  publisher={Oxford University Press},
DOI={ https://doi.org/10.1093/mnras/stz3223}
}

@article{PH69,
title = {Anomalous 52+ and 72+ states in 99Rh from the decay of 21 min 99Pd},
journal = {Nuclear Physics A},
volume = {135},
number = {1},
pages = {116-138},
year = {1969},
issn = {0375-9474},
doi = {https://doi.org/10.1016/0375-9474(69)90151-1},
url = {https://www.sciencedirect.com/science/article/pii/0375947469901511},
author = {M.E. Phelps and D.G. Sarantites}
}

@article{Kr25,
  title = {Direct mass measurement of $^{93}\mathrm{Pd}$ and implications for the isomer structures in $^{94}\mathrm{Ag}$: Tracing the two-proton decay branch},
  author = {Kripk\'o-Koncz, Gabriella and Pla\ss{}, Wolfgang R. and Dedes, Irene and Dao, Duy Duc and Dickel, Timo and Hornung, Christine and Amanbayev, Daler and Andr\'es, Samuel Ayet San and Beck, S\"onke and Bergmann, Julian and Blazhev, Andrey and Dudek, Jerzy and Geissel, Hans and Haettner, Emma and Kalantar-Nayestanaki, Nasser and Mardor, Israel and Miskun, Ivan and Mollaebrahimi, Ali and Mougeot, Xavier and Mukha, Ivan and Nowacki, Fr\'ed\'eric and Scheidenberger, Christoph and \"Ayst\"o, Juha and Bagchi, Soumya and Balabanski, Dimiter L. and Baran, Andrzej and Bren\ifmmode \check{c}\else \v{c}\fi{}i\ifmmode \check{c}\else \v{c}\fi{}, \ifmmode \check{Z}\else \v{Z}\fi{}iga and Charviakova, Volha and Constantin, Paul and Dehghan, Masoumeh and Gaamouci, Abdelghafar and Ge, Zhuang and G\'orska, Magdalena and Gr\"of, Lizzy and Hall, Oscar and Harakeh, Muhsin N. and Hucka, Jan-Paul and Kankainen, Anu and Kn\"obel, Ronja and Kostyleva, Daria A. and Kurkova, Natalia and Kuzminchuk, Natalia and Nichita, Dragos and Patyk, Zygmunt and Pietri, Stephane and Purushothaman, Sivaji and Reiter, Moritz Pascal and Reponen, Mikael and Roesch, Heidi and Sp\ifmmode \u{a}\else \u{a}\fi{}taru, Anamaria and Stanic, Goran and State, Alexandru and Tanaka, Yoshiki K. and Vencelj, Matja\ifmmode \check{z}\else \v{z}\fi{} and Weick, Helmut and Yang, Jie and Yavor, Mikhail I. and Zhao, Jianwei},
  collaboration = {for the Super-FRS Experiment Collaboration},
  journal = {Phys. Rev. Res.},
  volume = {7},
  issue = {4},
  pages = {L042022},
  numpages = {9},
  year = {2025},
  month = {Oct},
  publisher = {American Physical Society},
  doi = {10.1103/mhhn-kmgx},
  url = {https://link.aps.org/doi/10.1103/mhhn-kmgx}
}

@article{Wa11,
doi = {10.1088/0004-637X/729/1/46},
url = {https://doi.org/10.1088/0004-637X/729/1/46},
year = {2011},
month = {feb},
publisher = {The American Astronomical Society},
volume = {729},
number = {1},
pages = {46},
author = {Wanajo, Shinya and Janka, Hans-Thomas and Kubono, Shigeru},
title = {UNCERTAINTIES IN THE νp-PROCESS: SUPERNOVA DYNAMICS VERSUS NUCLEAR PHYSICS},
journal = {The Astrophysical Journal}
}

@article{Sa22,
doi = {10.3847/1538-4357/ac34f8},
url = {https://doi.org/10.3847/1538-4357/ac34f8},
year = {2022},
month = {jan},
publisher = {The American Astronomical Society},
volume = {924},
number = {1},
pages = {29},
author = {Sasaki, Hirokazu and Yamazaki, Yuta and Kajino, Toshitaka and Kusakabe, Motohiko and Hayakawa, Takehito and Cheoun, Myung-Ki and Ko, Heamin and Mathews, Grant J.},
title = {Impact of Hypernova νp-process Nucleosynthesis on the Galactic Chemical Evolution of Mo and Ru},
journal = {The Astrophysical Journal}
}

\end{document}